\documentclass[10pt, letterpaper, conference]{IEEEtran}

\usepackage[american]{babel} 
\usepackage{csquotes} 
\usepackage[babel=true]{microtype} 

\usepackage{booktabs}

\usepackage{graphicx}
\addto\captionsamerican{} 

\usepackage{mathtools}
\usepackage{stix}

\usepackage[framed, amsthm, hyperref]{ntheorem}
\theoremstyle{definition}
\newtheorem{definition}{Definition}

\usepackage{balance}

\usepackage{cite}

\newcommand{\Hair}{\ifmmode\mskip1mu\else\kern0.08em\fi}

\title{A Model-Based Approach to Security Analysis\\ for Cyber-Physical Systems}

\IEEEoverridecommandlockouts

\IEEEpubid{\makebox[\columnwidth]{978-1-5386-3664-0/18/\$31.00 \copyright 2018 IEEE \hfill} \hspace{\columnsep}\makebox[\columnwidth]{ }}

\author{ \IEEEauthorblockN {
	Georgios Bakirtzis,\IEEEauthorrefmark{1}\IEEEauthorrefmark{2}\IEEEauthorrefmark{3}
	Bryan T. Carter,\IEEEauthorrefmark{2}\IEEEauthorrefmark{4}
	Carl R. Elks,\IEEEauthorrefmark{1}\IEEEauthorrefmark{5} and
	Cody H. Fleming\IEEEauthorrefmark{2}\IEEEauthorrefmark{6}
        \thanks{This research is based upon work supported by the Systems Engineering
          Research Center under Award No. 2017-RT-172.}}
\IEEEauthorblockA {
	\IEEEauthorrefmark{1}Dependable Cyber-Physical Systems Lab, Electrical \& Computer Engineering, VCU, Richmond, VA USA\\
	\IEEEauthorrefmark{2}Coordinated Systems Lab, Systems \& Information Engineering, UVA, Charlottesville, VA USA\\
	\IEEEauthorrefmark{3}bakirtzisg@ieee.org
	\IEEEauthorrefmark{4}bcarter@virginia.edu
	\IEEEauthorrefmark{5}crelks@vcu.edu
	\IEEEauthorrefmark{6}fleming@virginia.edu
    }
}

\date{}

\begin{document}
\maketitle

\begin{abstract}
Evaluating the security of cyber-physical systems throughout their life cycle is necessary to assure that they can be deployed and operated in safety-critical applications, such as infrastructure, military, and transportation. Most safety and security decisions that can have major effects on mitigation strategy options after deployment are made early in the system's life cycle. To allow for a vulnerability analysis before deployment, a sufficient well-formed model has to be constructed. To construct such a model we produce a taxonomy of attributes; that is, a generalized schema for system attributes. This schema captures the necessary specificity that characterizes a possible real system and can also map to the attack vector space associated with the model's attributes. In this way, we can match possible attack vectors and provide architectural mitigation at the design phase. We present a model of a flight control system encoded in the Systems Modeling Language, commonly known as SysML, but also show agnosticism with respect to the modeling language or tool used.
\end{abstract}

\section{Introduction}
A major challenge in Cyber-Physical Systems (CPS)~\cite{national_interim_2015} is the assessment of the system's security posture at the early stages of its life cycle. In the defense community, it has been estimated that 70-80\% of the decisions affecting safety and security are made in the early concept development stages of a project~\cite{frola_system_1984,kutz_mechanical_2015,corbett_design_1986,saravi_estimating_2008}. Therefore, it is advantageous for this assessment to take place before lines of code are written and designs are finalized. To allow for security analysis at the design phase, a system model has to be constructed, and that model must reasonably characterize a system and be sufficiently detailed to enable matching attack vectors mined from databases. Matching possible attack vectors to the system model facilitates detection of possible security vulnerabilities in timely fashion. One can then design systems that are secure by design instead of potentially having to add \textit{bolt-on} security features later in the process, an approach that can be prohibitively expensive and limited in its mitigation options. Consequently, employing a model reduces costs and highlights the importance of security as part of the design process of CPS. We propose a model encoded in the Systems Modeling Language (SysML), a graphical object oriented modeling language~\cite{hause_sysml_2006}.

Our justification for using a solution based on SysML is twofold: it facilitates the implementation of any changes to the design of CPS and is a  tool familiar to and often used by systems engineers. However, as we do not want to limit the core ideas to SysML, we show that the model is transformable to a graph structure, thus demonstrating that it is agnostic to the particular modeling language or tool used.

To accurately characterize the system with respect to its associated security audit, the model requires a taxonomic scheme consisting of predefined categories. This schema is necessary to describe a potentially realizable or realized system. By extension the model, then, represents a \textit{real} system without necessarily requiring one to be built. To develop this schema, we investigate attack vector databases and examine their entries and their intra and inter connections. To represent the reality of the system, the chosen categories and their corresponding attributes need to correctly capture the hardware and software elements controlling the CPS as well as the interactions in the system model. The categories must be chosen in such a way that the model has sufficient fidelity and can be used to find attack vectors from open databases.

Since the model drives the vulnerability analysis, it needs to reflect relevant cyber-oriented information required to match possible attack vectors within the aforementioned categories. This additional information is encoded in the attributes of a subsystem and normally comes from preexisting design documentation, subject matter expertise, and requirements documentation. The amount and level of specificity that needs to be embodied in the attributes of each component depends on whether it matches the natural language that describes the attack vectors within the vulnerability databases. Therefore, the attributes need to represent the system's hardware and software composition so that they match possible entries in the databases. Only then can a system model become attackable, enabling us to infer possible vulnerabilities in the system's architecture and propose preemption and mitigation strategies.

Construction of a model with the characteristics described above gives rise to two main challenges. The first is the complexity in creating a complete schema that is able to capture---through the attributes---the necessary detail to represent the cyber aspects of an actual system. This first challenge also needs to take into account the importance of reasonably allowing the maintenance of any given CPS model that contains both cyber attributes and non-cyber attributes. The second is the difficulty in coordinating the available vulnerability data, the hierarchy within the databases that contain that data, and finally, the way in which the information is captured in their entries. 

\subsection{The Challenge of Capturing a Sufficient Set of Attributes}
\begin{figure}[!t]
\includegraphics[width=0.45\textwidth]{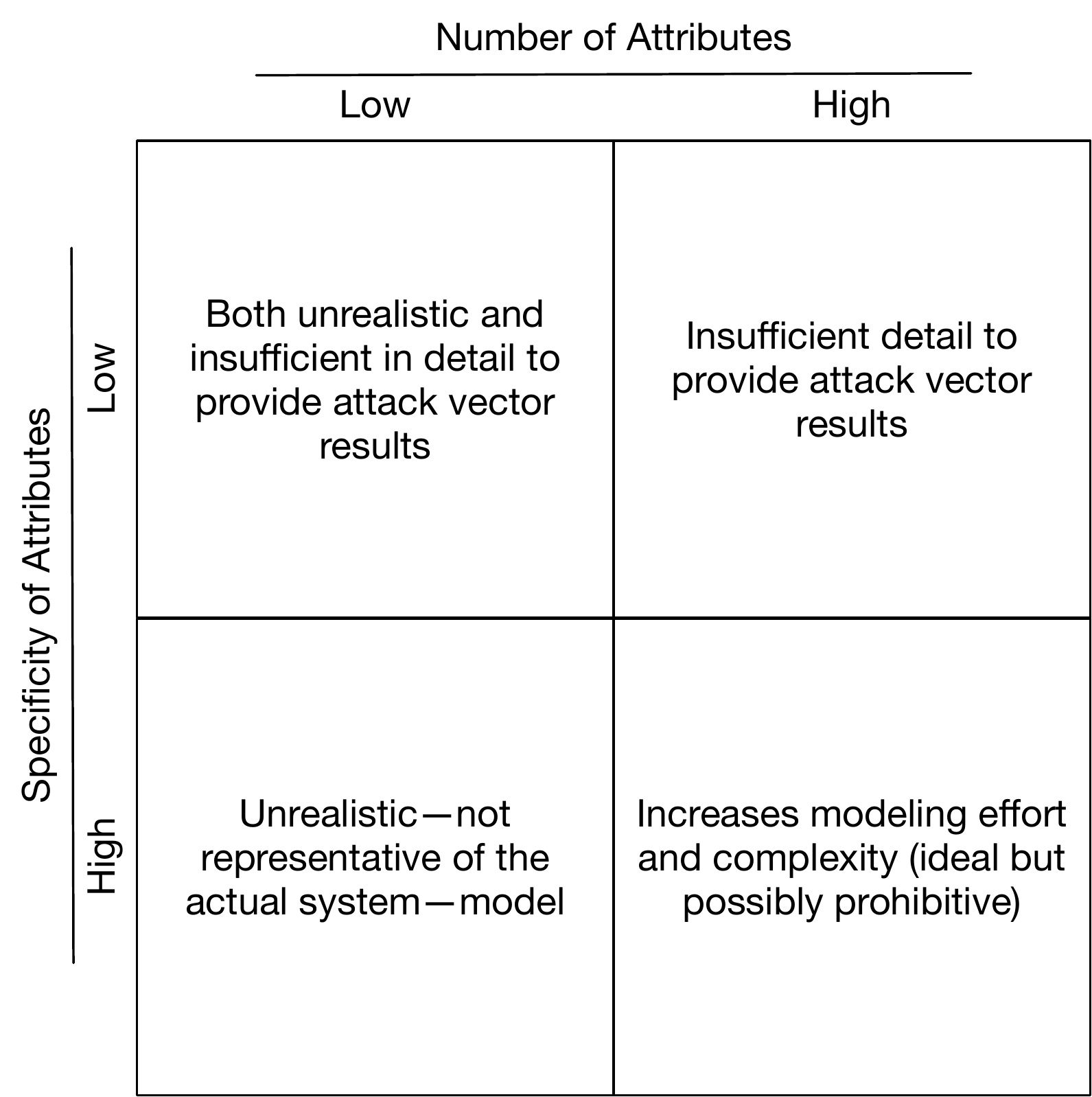}
\caption{The fidelity of the attributes describing a CPS has to achieve a balance between the amount of information that is captured in the model and the difficulty of producing it. At the extremes, the attributes can either be uninformative and incomplete or too detailed, requiring a prohibitive amount of modeling effort. Part of the challenge to being model sufficient is producing a sound, well-formed model for vulnerability assessment that can be practically utilized by systems engineers.}
\label{fig:grid}
\end{figure}

Indeed, there are tradeoffs between specificity and the number of attributes we can impose on a modeling methodology before it becomes impractical. This tradeoff can be illustrated by considering the extremities of model fidelity (Fig.~\ref{fig:grid}). On the one side of the spectrum (Low-Low) the model does not correctly characterize an actual system and its potential vulnerability to existing threats, while on the other side of the spectrum (High-High) the model requires a prohibitive amount of modeling effort.

Being model-driven in the context of cybersecurity has an additional requirement beyond ensuring fidelity to the real system. It also requires integrating the attributes taken from a design specification documentation, such that an analyst can match possible attack patterns to the model. This challenge is partially solved by understanding and using several repositories of attacks and vulnerabilities (Section \ref{sec:attack-vectors}). However, it still does not guarantee that possible query words---word lists that are used to associate potential vulnerabilities with a given set of attributes---will produce attack vectors applicable to that subsystem. Therefore, we need to capture the design documentation only to a point where there is agreement between two perspectives; (i) fidelity to the system's behavior and structure and (ii) the system's corresponding attack vectors.

\subsection{Challenge of Understanding Diverse Vulnerability Data}
Unfortunately, there is no single repository that contains all possible exploitation techniques or vulnerabilities that can apply to complex systems like CPS. If there was one, it would have to span across several domains---e.g., embedded devices, networks, humans in the loop---and have multiple levels of specificity for each entry in order to match every element in the system model. We leverage several databases with the goal of addressing all domains and at different levels of specificity, thus leading to new insights about the system's security posture. We then need to understand the different possible repositories, what information they contain, how they capture that information in natural language and what the targeted scope of each database is. By gaining that understanding, we extend the schema to incorporate data from diverse resources and ensure a thorough and rigorous security assessment.

Finally, the level of abstraction of each repository has to match the attributes in the model. At the very least, this requires a database that matches the attributes in the model. Ideally, however, there would be several databases that provide layers of abstraction. This is the case with a set of databases that are hierarchical and interconnected, e.g., CAPEC,\Hair\footnote{Common Attack Pattern Enumeration \& Classification, capec.mitre.org} CWE,\Hair\footnote{Common Weakness Enumeration, cwe.mitre.org} CVE.\Hair\footnote{Common Vulnerabilities and Exposures, cve.mitre.org} 
For this set of databases, CVE can match to the model attributes (since its vulnerability entries resides at a lower level of abstraction than the other two) and then be used to infer possible classes of attacks by mapping CVE entries to their corresponding, higher-level, CWE and/or CAPEC entries. The utility of this hierarchy is also important when reasoning about the threat space associated with the system. By abstracting the individual exploited vulnerabilities to more general weaknesses and/or patterns, this approach reduces the amount of information an analyst has to parse and reason with when they inspect a complex system model for applicable attacks.

\subsection{State of the Art}
Model-based techniques for assessing cybersecurity have been at the forefront of recent research in CPS. These traditionally stem from dependability and safety analysis. Nicol et al.~\cite{nicol_model-based_2004} have stated the need for model-based methods for assessing security that come from the general area of dependability. Further, Chen et al.~\cite{chen_go_2013} have proposed a model-based graph oriented analysis technique for assessing a system for acceptable safety based on a workflow. Kopetz~\cite{kopetz_real_2011} presents the notion of categories of interfaces to model real-time systems. Davis et al.~\cite{davis_cyber-physical_2015} present a framework that extends the notion of dependability to include possible security violation for the power grid that utilizes state estimation and is evaluated in a simulated model of the power grid. More recently, Brunner et al.~\cite{brunner_towards_2017} proposed a combined model for safety and security that is based on Unified Modeling Language (UML) diagrams. These models reside in a higher level of abstraction than our proposed model; they do not contain the structure of the system, do not consider evidence-based security assessment, assessment based on previous reported vulnerability data, and the evaluation targets certification of policy standards.

In general, little work has been done to determine whether a model contains the necessary attributes stemming from the design documents and encoded through a structural model to be used in evidence-based cyber-vulnerability assessment. Even less work has been done in targeting the model sufficiency of CPS and how that is used at the early stages of the design process. For the cybersecurity assessment of CPS, no standard rule-of-thumb, or otherwise generally accepted procedure has been established.

\subsection{Contributions of this Research}
The central contribution of this paper is the characterization and definition of the problem of agreement between system model and historic attack vector databases and tackling the difficulty of matching the two when it comes to CPS. Therefore, this research presents solutions to the two main challenges outlined above. First, it shows how to capture the relevant information within those schema categories based on design documentation that preexists the model. Second, it demonstrates how to appropriately handle historic data to identify possible vulnerabilities in the system's design. Both challenges can be solved by methodically constructing a model through a predefined taxonomic scheme. Toward those objectives, the model is built based on the following insights:

\begin{enumerate}
\item a necessary understanding of historic information is needed to match against a system description;
\item a system description needs to be evaluated as being realistic and it is characterized by attributes to an extent that can match possible attack vectors; and
\item agnosticism toward modeling language or tool.
\end{enumerate}

This paper focuses on model sufficiency with respect to vulnerabilities, with the explicit recognition that these vulnerabilities can give rise to unsafe or undesired behavior in the overall, coupled CPS. Moreover, this paper demonstrates how vulnerabilities propagate to physical system behavior but it is outside the defined scope to analyze the physical behavior and/or determine whether it is (un)safe or (un)desirable.

To assess the sufficiency of the model we present a model of a generic Flight Control System (FCS) and assess the possible security violations of two system components using open vulnerability databases.

\section{A Taxonomic Scheme for CPS Attributes}

Our main objective is to construct a general purpose taxonomic scheme that can be used to characterize the cyber components of a CPS and their interactions, for the purpose of relating to attack vectors. To this end, we use preexisting design specification documentation to describe the attributes of the cyber components and encode this information in the model. To methodically achieve this we first present definitions for cyber component, attack vector, cyber attribute, evidence, and taxonomic scheme. These definitions provide common ground on the relatively generic term ``cyber,'' which is used in several contexts and, therefore, can hold different meanings. 

\subsection{Primitives}

Our model of CPS makes a distinction between cyber and physical components. Caution is necessary because the components of a CPS can reside in between the cyber and physical realms but their behavior and form can be described by either. This paper is intended only to identify the minimum set of attributes necessary to assess a CPS's cybersecurity posture. 

\begin{definition}
A \textsc{cyber component} of a cyber-physical system is any device that is programmable and whose associated computation observes or controls physical quantities, e.g., velocity, altitude, etc.
\end{definition}

\begin{definition}
An \textsc{attack vector} is a specific description of an attack on a given subsystem. It presents the features of the exploited vulnerability, the privilege access level required for an attacker to perform the attack, and the steps to perform it.
\end{definition}

\begin{definition}
A \textsc{cyber attribute} defining a subsystem of a cyber-physical system represents possible specification of behavior, form, or structure. Hence, the set of attributes produce an architectural definition of the subsystem and represent the part(s) of the model that maps to possible attack vectors.
\end{definition}

\begin{definition}
\textsc{Evidence} is all instances of historic vulnerability data---attack vectors---that can be mapped to a cyber component through any cyber attribute or a combination of cyber attributes.
\end{definition}

\begin{definition}
A \textsc{taxonomic scheme}, or schema, is a discrete set of categories that can capture the structure of any cyber component in a cyber-physical system to produce evidence.
\end{definition}

Following the above definitions, the taxonomic scheme should inform the following questions:

\begin{enumerate}
\item What is the subsystem?
\item How is it implemented?
\item Who does it talk to?
\item Why is it there?
\end{enumerate}

The first question informs the model about the identity of the subsystem. The second question provides the design details, used by the model to characterize a possible real subsystem. The third question identifies the required interactions between the subsystem, ensuring that the composition of the full system can provide its expected service---an important aspect of CPS as indicated throughout the literature~\cite{sun_architecture_2016}. The fourth and last question addresses the function of the subsystem and, therefore\ informs about its relative criticality to the overall behavior of CPS.

Answering the above questions sufficiently and constructing a taxonomic scheme directed by them allows us to base our analysis in methodical reasoning. Furthermore, by utilizing methodical reasoning we are able to view the threat spaces of a system holistically and take into account that a single segregated component acts differently than when it coordinates with other subsystems to compose a CPS. By using this approach, one is better informed about the overall system security posture than by analyzing just the system's components individually.

\subsection{Realization of the Taxonomic Scheme}

Using the above questions as a guide and being aware of the intrinsic structure and specificity contained in open vulnerability databases, the following taxonomic scheme composes the structure of any CPS and can assist in producing evidence: 

\begin{itemize}
\item \textbf{Operating System.} Since CPS is composed of hardware and software, it is important to be aware of the system software hosted, such as Real-Time Operating Systems (RTOS), executives, debuggers. In some cases, the embedded devices may be programmed without an operating system, a common term for which is ``programmed on bare metal.'' Knowing this information---that is, CPS running an embedded operating system---informs on possible vulnerabilities, e.g., bugs in the Linux kernel.

\item \textbf{Device Name.} The specific naming of a subsystem can assist in finding device-specific vulnerabilities.

\item \textbf{Hardware.} The decomposition of the specific device to its possible exploitable hardware elements, e.g., what chipset is on board.

\item \textbf{Firmware.} The possible firmware and corresponding drivers necessary to run the device.

\item \textbf{Software.} In the event that the subsystem runs an RTOS it is of importance to know the possible software that is installed and can potentially introduce further vulnerabilities.

\item \textbf{Communication.} Any cyber or physical interaction the CPS must implement in to provide its expected service.

\item \textbf{Entry Points.} All possible accessible entry points to the system. This attribute allows us to filter components that are part of the attack surface.
\end{itemize}

\subsection{Attributes}

In accordance with the taxonomic scheme above, a minimum set of attributes is presented in Table~\ref{tab:attributes}. The example comes from an NMEA GPS and its corresponding attributes stem from design documents and are refined using data sheets. Further refinement of those attributes is allowed. This dissemination is mostly based on the information provided in the design requirements documents but can also include information provided by subject matter experts.

\begin{table}[!t]
\renewcommand{\arraystretch}{1.3}
\caption{A GPS example of the minimum set of attributes necessary to create a sound, well-formed model of CPS. These attributes need to be used for any given subsystem that is pertinent to its expected service. The matching of attack vectors derives from the attributes specified in this table.}
\label{tab:attributes}
\centering
\begin{tabular}{ll}
\multicolumn{2}{c}{NMEA GPS} \\
\toprule
Category& Attributes\\
\midrule
Operating system & Bare metal\\
Device Name& Adafruit Ultimate GPS\\
Hardware & Mediatek MTK 3339 chipset\\
Firmware & Communication protocol drivers\\
Software & \(\emptyset\)\\
Communication & I2C, RS232, UART, RF\\
Entry Points & RF\\
\bottomrule
\end{tabular}
\end{table}

\section{SysML Model}

\begin{figure*}[!t]
\includegraphics[width=1.0\textwidth]{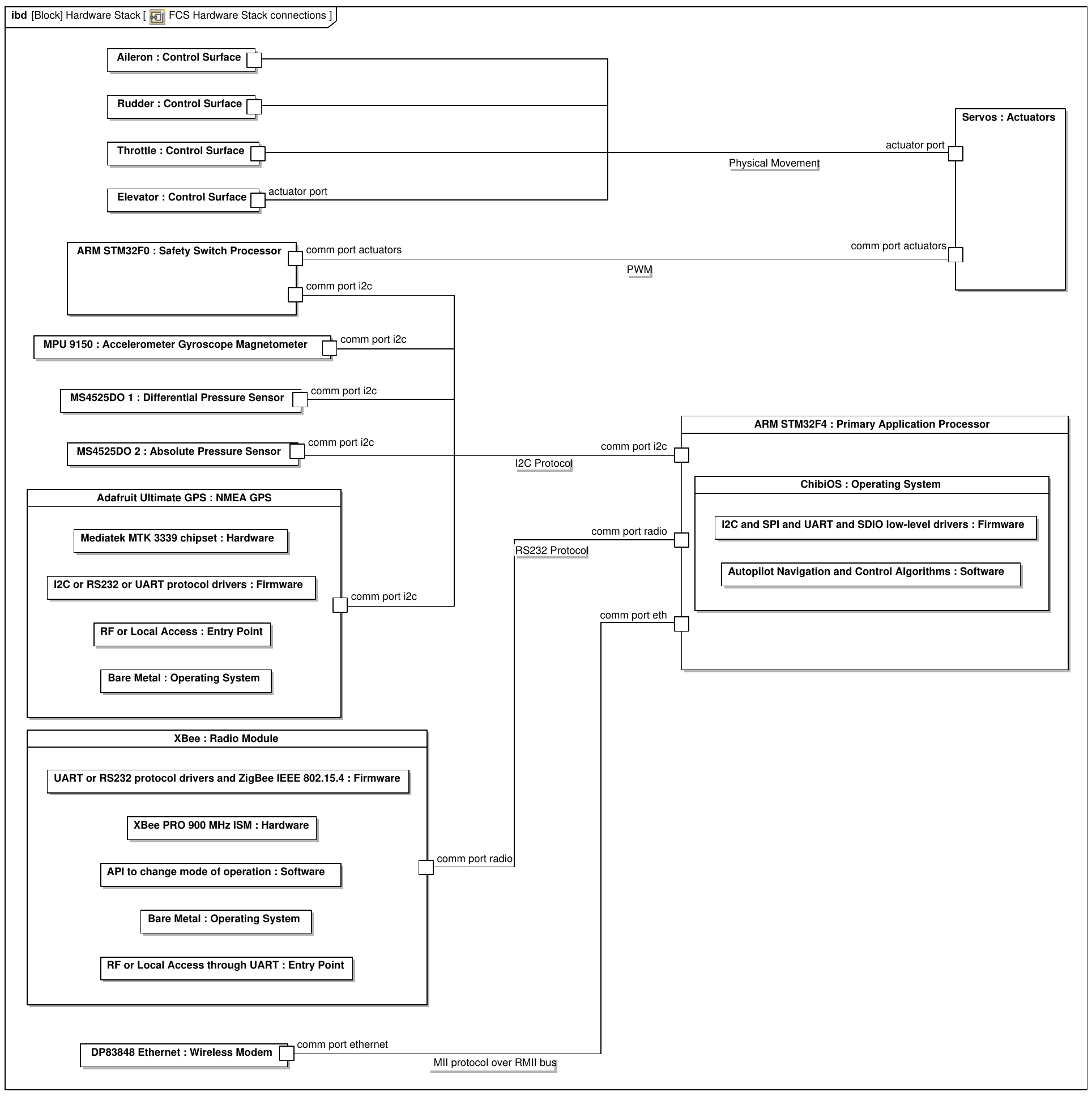}
\caption{The Internal Block Diagram (IBD) SysML model of the Flight Control System (FCS) with all the necessary attributes to characterize the primary application processor, the NMEA GPS, and the radio module. This model can inform about possible architectural changes at the early stages of design to avoid using components that have reported vulnerabilities or clearly mitigate against these vulnerabilities through better informed requirements. As such, we are able to construct systems that are secure by design.}
\label{fig:ibd}
\end{figure*}

Consider, for example, an Unmanned Aerial System (UAS) that is used to assist in search and rescue operations where minutes (or even seconds) count. In this domain, losing a UAS is certainly going to risk longer mission times and can potentially lead to an unsuccessful search and rescue operation. For these types of safety-critical missions, it is essential to assess the security posture of a given FCS design, so that a threat actor cannot interfere with safety-critical operations. For that reason we construct and evaluate an FCS model for possible security violation in an evidence-based fashion.

In the most general sense an FCS implements all the capabilities and control surfaces needed to operate an autopilot~\cite{ward_modular_2014,ward_design_2014}. This autopilot is used to fly UAS and is usually controlled through a ground control station. An FCS is composed of several compromisable cyber subsystems whose intended function is to modify and control physical parameters, e.g., control of engine throttle for speed, changing the direction based on a navigational goal by controlling the aileron, etc. By this definition, an FCS is a CPS and prime example of analysis given its ubiquity and utility in many domain areas. An FCS is safety-critical system, in the sense that if there is a hazard, either artificial through a cyber attack or by natural faults, it can cause severe consequences. 

The model of the FCS is encoded in an Internal Block Diagram (IBD) (Fig.~\ref{fig:ibd}). An IBD representation is used to define a system's structure. Traditionally, this model contains only generic representations of subsystems, e.g., power system, engine sensor, magnetometer and general information about flows of information, e.g., energy, command, sensor measurement. The taxonomic scheme described in the previous section adds specific implementation information that assists threat analysts in finding possible vulnerabilities and, consequently, associated attack vectors. This specific implementation information is encoded using part properties for component attributes, e.g., what type of GPS, and connectors for interaction attributes, e.g., using the I2C protocol.

Part properties encode further attribute information in a manner that is easy to parse by threat analysts. Part properties, as the name implies, are attributes that can further characterize an IBD block and take the form of \(\langle \text{part~name} \rangle
~:~\langle \text{type} \rangle\). Additionally, they can be decomposed to further part properties to make general categories of attributes---this can be seen by the decomposition of the \textit{Operating System} type in the \textit{Primary Application Processor} to further part properties (Fig.~\ref{fig:ibd}). Moreover, part properties can construct a new IBD by themselves to define the connectivity between components in a collection of part properties. This is useful in more complex systems where the connections might have different levels of abstraction. Part properties define the structure and composition of the system. 

Finally, connector types define source and target relationships as well as the type of interaction, digital protocols, analog inputs, or possibly physical actions. For example, the connection between the \textit{Safety Switch Processor} is digital and uses Pulse Width Modulation (PWM) commands to move the servos (Fig.~\ref{fig:ibd}). The servos then provide the physical pull to either open the throttle further---to gain velocity---or change the direction of movement by controlling the aileron.

\begin{figure*}[!t]
\includegraphics[width=1.0\textwidth]{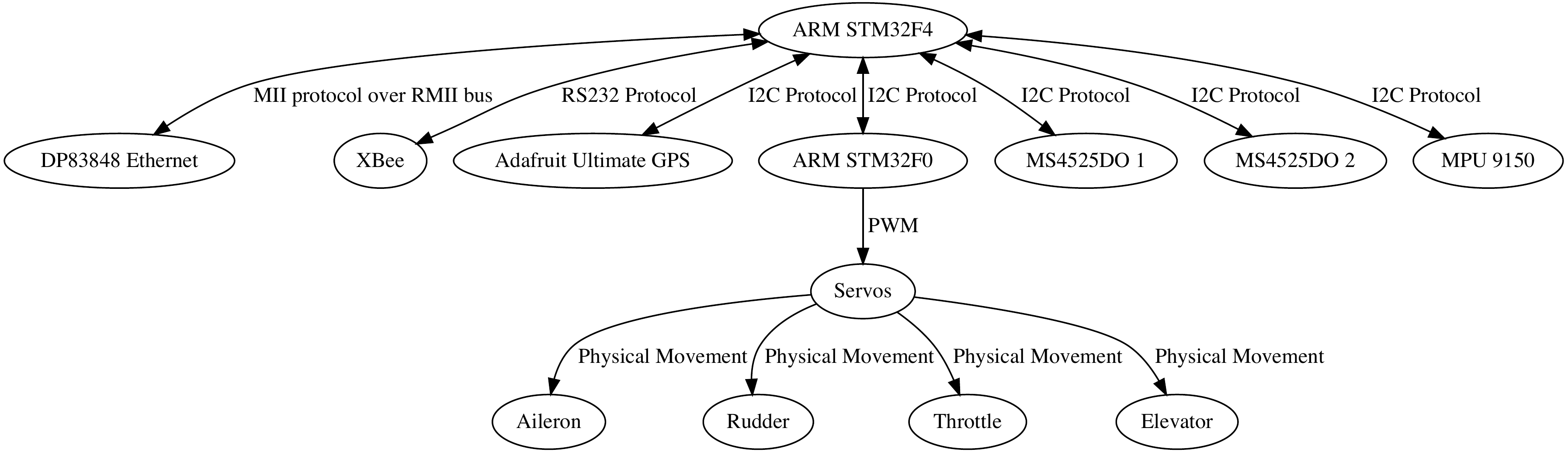}
\caption{The Internal Block Diagram (IBD) (Fig.~\ref{fig:ibd}) of the Flight Control System (FCS) maps directly to a graph structure (the part property type is omitted for visualization purposes). This allows us to extract the elements in an internal block diagram without loss of any information vital for cybersecurity assessment. The graph structure can be further input to other analysis techniques and provide a schema for the topological definition of a Cyber-Physical System (CPS). In this instance the vertex attributes can be accessed through the GraphML specification even though they are not visualized here.}
\label{fig:graph}
\end{figure*}

\section{Model Transformation}

A model transformation is used so that the modeling methodology presented in this paper is agnostic to specific modeling tools and languages.
To achieve this transformation, we construct a generic formalism for SysML IBD based on graph structures. The formalism is used as a basis for a tool which extracts the information from the IBD and encodes it in GraphML. GraphML is based on XML and consists of the de facto schema for sharing graphs~\cite{brandes_graph_2013}. Transforming SysML models to a graph should allow, in the future, the model to be analyzed with a variety techniques. This is potentially beneficial because SysML has limited verification capabilities and requires a specific modeling methodology to produce validated results. The following formalism assures that all SysML information and their corresponding properties are not lost in the transformation to GraphML. 

\subsection{Formal Semantics of Internal Block Diagrams}

Computing networks are typically reasoned about using graph structures, where the vertices represent assets and the edges represent their immediate connections. This is no different for CPS~\cite{weaver_toward_2013}. This paper extends the definition of assets to encompass every subsystem in a CPS, which comprises of more than the computing systems, including but not limited to imagery payload, actuators, sensors and their data links to the computing systems. We formalize these definitions, which are initially encoded in a SysML IBD, using standard graph notation below.

\begin{definition}
An \textsc{internal block diagram} is a graph \(G \coloneqq (V, P, \textit{src}, \textit{tgt}, \mathcal{A})\), where \(V\) is the set of vertices of \(G\); \(P\) is the set of ports of \(G\); \(\textit{src}, \textit{tgt}: P \rightarrow V\) functions source and target for \(G\) respectively, and \(\mathcal{A}\) is the set of attributes of \(G\). \(V\) represents the components of a cyber-physical system, \(P\) the inputs and outputs corresponding to the components, \(\textit{src}, \textit{tgt}\) the directionality of the possible cyber or physical interactions between components, and \(\mathcal{A}\) the associated descriptors for a given vertex or connection.
\end{definition}

An example transformed system graph is depicted in Fig.~\ref{fig:graph} where not all information is necessarily visualized but is encoded in the GraphML format and can be accessed programatically.

\begin{definition}
A \textsc{part property} in an internal block diagram is a function \(\textit{attr}_v: V \rightarrow \mathcal{A}_{v}\), where \(V\) is the set of vertices of \(G\) and \(\mathcal{A}_{v}\) the set of vertex attributes, such that \(\mathcal{A}_{v} \subseteq \mathcal{A}\).
\end{definition}

Consider a single mapping for the attributes of the NMEA GPS, (Table~\ref{tab:attributes} and Fig.~\ref{fig:ibd}):
\begin{equation*}
\begin{split}
\textit{attr}_v(\text{GPS}) \mapsto \{\text{Bare Metal}, \text{Adafruit Ultimate GPS},\\
\text{Mediatek MTK 3339 chipset}, \ldots, \text{RF}\}.
\end{split}
\end{equation*}

\begin{definition}
A \textsc{connector} in an internal block diagram is a function \(\textit{attr}_p: P \rightarrow \mathcal{A}_p\), where \(P\) is the set of ports of \(G\) and \(\mathcal{A}_{p}\) the set of port attributes, such that \(\mathcal{A}_p \subseteq \mathcal{A}\).
\end{definition}

For example, the tuple of vertices below passed into the port attribute function will provide the edge-specific attributes for that tuple: \(\textit{attr}_p(\{\text{GPS}, \text{STM32F4}\}) \mapsto \{\text{I2C Protocol}\}\).

Given the above definitions, we transform the SysML model to the neutral GraphML format and can further use it as input to other analysis techniques, including finding attack vectors.

\section{Matching Potential Attack Vectors\label{sec:attack-vectors}}

To evaluate the model we find and map applicable attack vectors for subsystems of the FCS model described above (Fig.~\ref{fig:ibd} and Fig.~\ref{fig:graph}). This analysis is based on analyzing open vulnerability databases to find possible attacks and provide possible design mitigation strategies. For example, given a component with an associated set of attack vectors, one can then assess the risk and potentially find another component that provides the same expected service without any reported vulnerabilities.

\subsection{Experimental Setting}

Toward the evaluation of the model's fidelity we find potential attack vectors for the subsystems of the FCS model using cve-search.\Hair\footnote{CVE-SEARCH PROJECT, cve-search.org (perma.cc/5J2M-VAGC)} This online database not only provides possible CVE entries that are applicable based on the query strings produced by the models attributes, but can also relate to higher levels of abstraction, e.g., CWE or CAPEC, to further understand the possible impact of a given attack.

We assume that a system is vulnerable if any single attack vector from the databases maps to any single attribute of the system model. For example, if an attack vector targets Operating System `A' with Driver `B', we consider it a vulnerability even if the model includes only `A' or `B'; it does not have to contain both, even if the attack pattern specifies them together. This assumption is reasonable because a large number of vulnerabilities that are reported have to do with systems that are popular and widely used, e.g., the Android operating system and corresponding drivers. This assumption allows for extrapolation from such (specific) reports to better understand the security posture of the model where, at the moment, no embedded system vulnerability database exists. 

Furthermore, it is uncommon for users to update their embedded devices, and this assumption still allows the analyst to take into account vulnerabilities that may have been fixed in newer versions of firmware or software.

Finally, we assume that a subsystem that has no reported attack vectors is more secure than one that does. This work does not focus on zero-days because there is currently no way---at the design phase---to assess zero-day vulnerability just by analyzing the architecture of the system. However, our approach also works using private or proprietary vulnerability databases. As long as there exist historic data on the given subsystem, this approach allows an analyst to identify them and better inform system designers. 
In attempting to violate the security posture of a system, a given intelligent threat actor will, most likely, use a set of techniques they are familiar with rather than come up with new techniques tailored to the individual system~\cite{allodi_work-averse_2017}.

\begin{figure}[!t]
\includegraphics[width=0.5\textwidth]{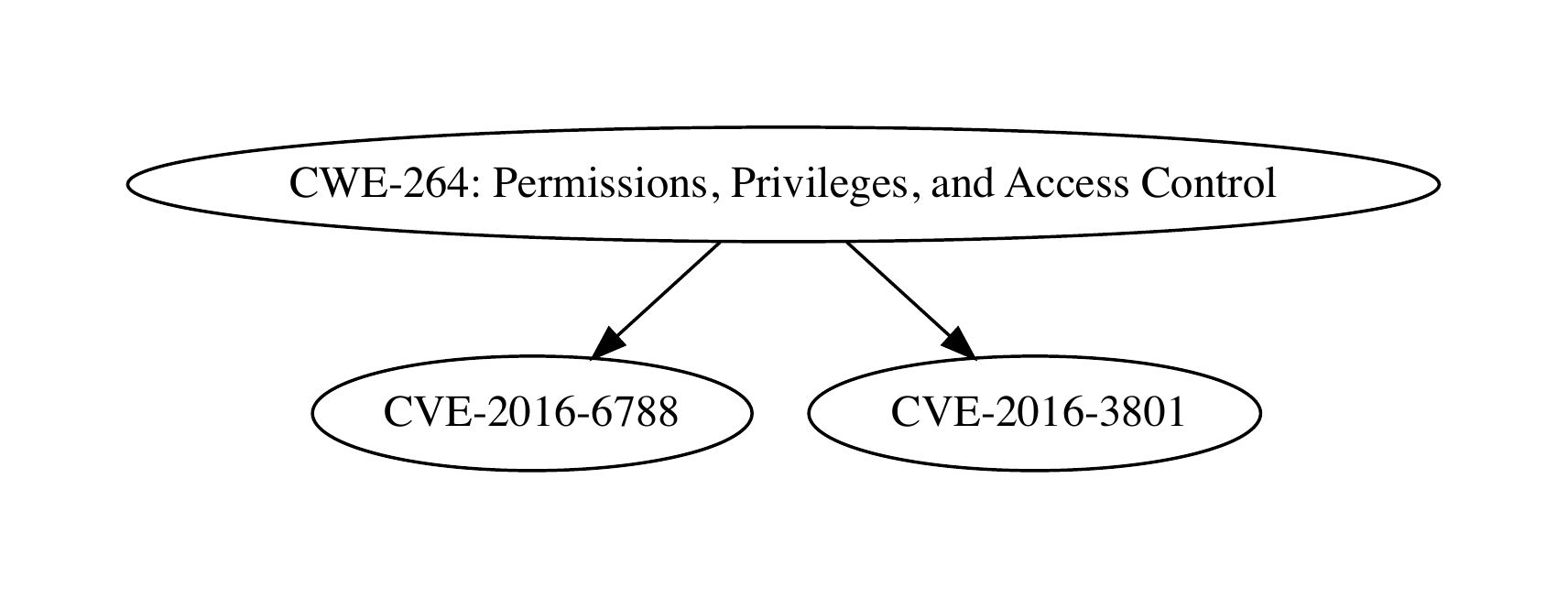}
\caption{The two possible Adafruit Ultimate GPS attack vectors found in CVE share the same CWE category. This can inform the designer about general issues with the specific, chosen subsystem and to possibly look at substitutes with no reported vulnerabilities. Otherwise, the designers might choose to clearly document this class of vulnerabilities and construct solid requirements to mitigate such possible violation of system assets.}
\label{fig:attack_vectors}
\end{figure}

\subsection{Results}

This section describes a security assessment of two components in the FCS that can cause full degradation of expected service through exploiting historic vulnerabilities. Specifically, this analysis focuses on (1) the NMEA GPS device, which is necessary to provide location data and  (2) the radio module, which is necessary to communicate with the ground control station or in the instance of manual control, the operator.

\begin{figure}[!t]
\includegraphics[width=0.5\textwidth]{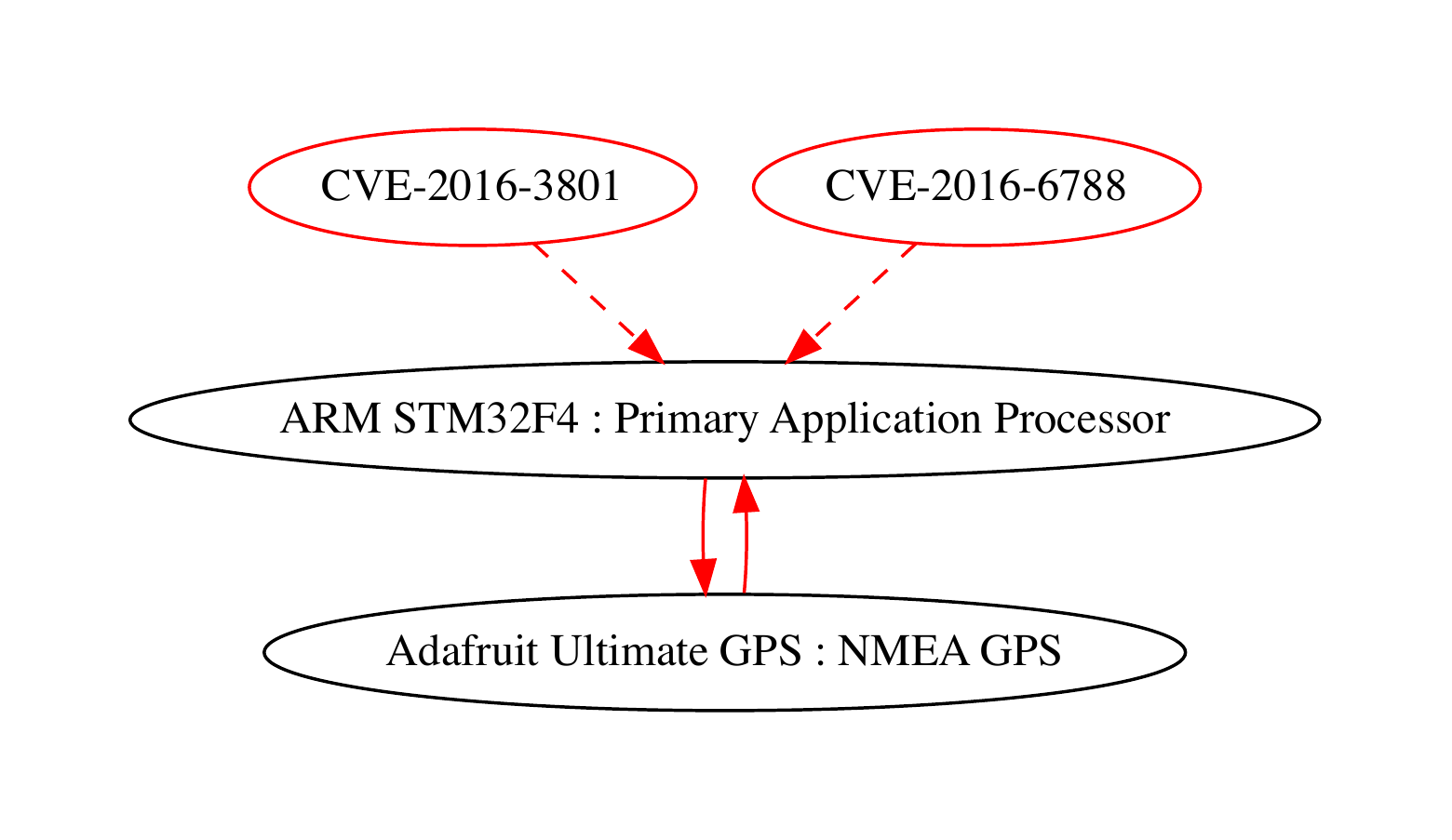}
\caption{An intelligent threat actor can potentially take advantage of the use of the Adafruit Ultimate GPS drivers and can completely violate the systems expected service by escalating their privileges by either using the attack vectors presented individually or for a higher impact in sequence (attack chain). The dashed red edges indicate a given attack step, while the red solid edges indicate violation of data exchange. Since one of the attacks can lead to arbitrary code execution it can violate any path from the microcontroller to the sensory systems, meaning that such an attack would cause full degradation of expected service.}
\label{fig:gps_attacks}
\end{figure}

\textbf{NMEA GPS.} \quad Given the specification of the Adafruit Ultimate GPS (Table~\ref{tab:attributes} and Fig.~\ref{fig:ibd}) we search through CVE to find two possible attack vectors (Fig.~\ref{fig:attack_vectors}). The first is CVE-2016-3801, which is a reported vulnerability specific to Android devices and targets the Mediatek GPS driver in the embedded operating system by crafting an application that can exploit the driver to \textit{gain} system privileges. Even if an Android device is not specifically in use, it is possible to misconfigure or program the FCS in such a way that the attack vectors are applicable, making this a threat one must account for when designing the system. The second is CVE-2016-6788, an attack that targets MediaTek I2C drivers and subsequently allows an attacker to \textit{elevate} their privileges and execute arbitrary code. This vulnerability was also reported for Android but, again, it can be applicable to the design of the FCS. Whilst both attacks take advantage of the GPS, they actually target the primary application processor (Fig.~\ref{fig:gps_attacks}) with possibly devastating effects to the system's expected service.

Further, the two vulnerabilities can form an attack chain by first using CVE-2016-3801 to gain access to the system, which would require an operator possibly accepting a request from the attacker, and then using CVE-2016-6788 to further elevate their privileges without having to go through operator input. It would be difficult to find these attacks without decomposing the NMEA GPS device down to its specific implementation including the chipset it is employing and the required firmware to operate. Hence, system designers would have been unaware of this possible attack chain and would have to add further security considerations for this component post-deployment, instead of switching it with another that has no historic reported vulnerabilities and, therefore, presumably less risk.

\textbf{Radio Module.} \quad Another part of the system that could be attacked is the XBee radio module, which requires drivers for the ZigBee IEEE 802.15.4 protocol. Its specification can be seen in the SysML diagram (Fig.~\ref{fig:ibd}). One of the possible attacks is described in CVE-2015-8732 and CVE-2015-6244 (these attacks have different bugs associated with them but result in the same effect), where an intelligent threat actor constructs packets that cause out-of-bound read and application crashes, resulting in a successful denial of service. Without radio communication the FCS would not be able to coordinate with the ground control station and it would go to a fail-safe mode, which could be detrimental to the mission it was planned to carry out. Knowledge of such attacks to the system's designers can lead to choosing a more robust, security wise, radio module for the implementation of the FCS.

\subsection{Discussion}

The aforementioned example results further illustrate the need for early detection of vulnerabilities. While this work has not focused on evaluating elements of the attack surface--that is, components exposed to intelligent threat actors that can act as entry points into the systems structure--this paper demonstrates that a model can assist in generating security by design. This approach generates evidence (the historic data from open databases) and in future work will be used to design the systems with hardware and software that is historically more secure at no additional cost, except those costs required for the security analysis. If architectural changes are not an option, it is still crucial to be aware of possible vulnerabilities and impose clear system requirements to preempt or mitigate against classes of attacks. 

Without the generalized taxonomic scheme and its generated specification for the system, we would not have been able to match evidence to subsystems. This could have led to insecure systems getting deployed for safety-critical applications, which in turn can cause hazardous behaviors or, in the worst-case scenario, controlled accidents by intelligent threat actors.

\section{Conclusions \& Beyond}

\balance

In this paper we have presented a framework that characterizes a characteristic set of attributes for each given subsystem in a CPS. These attributes construct well-formed models that are sufficiently detailed to allow for security posture evaluation of the system they specify. This framework is built on the examination of historic vulnerability data from databases, termed evidence, which apply to the system model based on those attributes. We have shown that this framework produces model sufficiency by mapping attack vectors for a possible NMEA GPS and radio module. While the method is agnostic to the modeling language, we represent the system in SysML, which is ubiquitous in systems engineering.

A future direction can use the findings of this research to automate the process of matching attack vectors. Towards this automation, we have shown the versatility of the method by transforming the SysML model to a generic graph metamodel using a standard graph schema based on XML, GraphML. A possible extension in this direction is to use the extracted system structure and apply techniques from computational linguistics to automatically produce attack vectors for each subsystems. This may allow threat analysts to assess the security posture of increasingly complex systems.

Finally, through this work we have provided an answer to a largely open question in the field of CPS. Namely, what level of design detail does the system model have to capture to allow for the evaluation of its security properties? The advantage of our model is that it views the system from its characteristic and identifiable attributes and does not depend on its function to provide a holistic view of its security posture.

\bibliographystyle{IEEEtran}
\bibliography{manuscript}

\begin{thebibliography}{10}
\providecommand{\url}[1]{#1}
\csname url@samestyle\endcsname
\providecommand{\newblock}{\relax}
\providecommand{\bibinfo}[2]{#2}
\providecommand{\BIBentrySTDinterwordspacing}{\spaceskip=0pt\relax}
\providecommand{\BIBentryALTinterwordstretchfactor}{4}
\providecommand{\BIBentryALTinterwordspacing}{\spaceskip=\fontdimen2\font plus
\BIBentryALTinterwordstretchfactor\fontdimen3\font minus
  \fontdimen4\font\relax}
\providecommand{\BIBforeignlanguage}[2]{{%
\expandafter\ifx\csname l@#1\endcsname\relax
\typeout{** WARNING: IEEEtran.bst: No hyphenation pattern has been}%
\typeout{** loaded for the language `#1'. Using the pattern for}%
\typeout{** the default language instead.}%
\else
\language=\csname l@#1\endcsname
\fi
#2}}
\providecommand{\BIBdecl}{\relax}
\BIBdecl

\bibitem{national_interim_2015}
{National Research Council}, ``Interim report on the 21st century
  cyber-physical systems education,'' National Academies of Sciences,
  Engineering, and Medicine, Tech. Rep., 2015.

\bibitem{frola_system_1984}
F.~Frola and C.~Miller, ``System safety in aircraft management,''
  \emph{Logistics Management Institute, Washington DC}, 1984.

\bibitem{kutz_mechanical_2015}
M.~Kutz, \emph{Mechanical Engineers' Handbook, Volume 2: Design,
  Instrumentation, and Controls}.\hskip 1em plus 0.5em minus 0.4em\relax John
  Wiley \& Sons, 2015.

\bibitem{corbett_design_1986}
J.~Corbett and J.~R. Crookall, ``Design for {Economic} {Manufacture},''
  \emph{CIRP Annals - Manufacturing Technology}, vol.~35, no.~1, pp. 93--97,
  Jan. 1986.

\bibitem{saravi_estimating_2008}
M.~Saravi, L.~Newnes, A.~R. Mileham, and Y.~M. Goh, ``Estimating cost at the
  conceptual design stage to optimize design in terms of performance and
  cost,'' in \emph{Proceedings of the 15th ISPE International Conference on
  Concurrent Engineering}.\hskip 1em plus 0.5em minus 0.4em\relax Springer,
  2008, pp. 123--130.

\bibitem{hause_sysml_2006}
M.~Hause, ``The {SysML} {Modelling} {Language},'' in \emph{Fifteenth {European}
  {Systems} {Engineering} {Conference}}, 2006.

\bibitem{nicol_model-based_2004}
D.~M. Nicol, W.~H. Sanders, and K.~S. Trivedi, ``Model-based evaluation: from
  dependability to security,'' \emph{{IEEE} Transactions on Dependable and
  Secure Computing}, vol.~1, no.~1, pp. 48--65, 2004.

\bibitem{chen_go_2013}
B.~Chen, Z.~Kalbarczyk, D.~M. Nicol, W.~H. Sanders, R.~Tan, W.~G. Temple, N.~O.
  Tippenhauer, A.~H. Vu, and D.~K.~Y. Yau, ``Go with the flow: Toward
  workflow-oriented security assessment,'' in \emph{Proceedings of New Security
  Paradigm Workshop ({NSPW})}, 2013.

\bibitem{kopetz_real_2011}
H.~Kopetz, \emph{Real-time systems: design principles for distributed embedded
  applications}.\hskip 1em plus 0.5em minus 0.4em\relax Springer Science \&
  Business Media, 2011.

\bibitem{davis_cyber-physical_2015}
K.~R. Davis, C.~M. Davis, S.~A. Zonouz, R.~B. Bobba, R.~Berthier, L.~Garcia,
  and P.~W. Sauer, ``A cyber-physical modeling and assessment framework for
  power grid infrastructures,'' \emph{{IEEE} Transactions on Smart Grid},
  vol.~6, no.~5, pp. 2464--2475, 2015.

\bibitem{brunner_towards_2017}
M.~Brunner, M.~Huber, C.~Sauerwein, and R.~Breu, ``Towards an integrated model
  for safety and security requirements of cyber-physical systems,'' in
  \emph{2017 {IEEE} International Conference on Software Quality, Reliability
  and Security Companion ({QRS}-C)}, 2017, pp. 334--340.

\bibitem{sun_architecture_2016}
C.~Sun, J.~Ma, and Q.~Yao, ``\BIBforeignlanguage{en}{On the architecture and
  development life cycle of secure cyber-physical systems},''
  \emph{\BIBforeignlanguage{en}{Journal of Communications and Information
  Networks}}, vol.~1, no.~4, pp. 1--21, Dec. 2016.

\bibitem{ward_modular_2014}
G.~L. Ward, G.~Bakirtzis, and R.~H. Klenke, ``A modular software platform for
  unmanned aerial vehicle autopilot systems,'' in \emph{52nd Aerospace Sciences
  Meeting}, ser. AIAA SciTech.\hskip 1em plus 0.5em minus 0.4em\relax American
  Institute of Aeronautics and Astronautics, Jan. 2014.

\bibitem{ward_design_2014}
G.~L. Ward, ``Design of a small form-factor flight control system,'' Master's
  thesis, Virginia Commonwealth University, 2014.

\bibitem{brandes_graph_2013}
U.~Brandes, M.~Eiglsperger, J.~Lerner, and C.~Pich, ``Graph markup language
  ({GraphML}),'' in \emph{Handbook of graph drawing visualization}, ser.
  Discrete mathematics and its applications, R.~Tamassia, Ed.\hskip 1em plus
  0.5em minus 0.4em\relax Boca Raton, FL: CRC Press, 2013, pp. 517--541.

\bibitem{weaver_toward_2013}
G.~A. Weaver, C.~Cheh, E.~J. Rogers, W.~H. Sanders, and D.~Gammel, ``Toward a
  cyber-physical topology language: Applications to {NERC} {CIP} audit,'' in
  \emph{Proceedings of the First {ACM} Workshop on Smart Energy Grid Security},
  ser. {SEGS} '13.\hskip 1em plus 0.5em minus 0.4em\relax {ACM}, 2013, pp.
  93--104.

\bibitem{allodi_work-averse_2017}
\BIBentryALTinterwordspacing
L.~Allodi, F.~Massacci, and J.~M. Williams, ``The work-averse cyber attacker
  model: Theory and evidence from two million attack signatures,'' Social
  Science Research Network, {SSRN} Scholarly Paper {ID} 2862299, 2017.
  [Online]. Available: \url{https://papers.ssrn.com/abstract=2862299}
\BIBentrySTDinterwordspacing

\end{thebibliography}
\end{document}